\documentclass[smallcondensed]{svjour3}     
\smartqed  
\usepackage{graphicx}
\begin{document}

\title{Fuzzy Characterization of Near-Earth-Asteroids
}


\author{Florian Freistetter}


\institute{Florian Freistetter \at
              Astrophysikalisches Institut und Universit\"atssternwarte \\
              Tel.: +123-45-678910\\
              Fax: +123-45-678910\\
              \email{florian@astro.uni-jena.de} }          

\date{Received: date / Accepted: date}

\maketitle

\begin{abstract}
Due to close encounters with the inner planets, Near-Earth-Asteroids (NEAs) can have very chaotic orbits. Because of this chaoticity, a statistical treatment of the dynamical properties of NEAs becomes difficult or even impossible. We propose a new way to classify NEAs by using methods from {\em Fuzzy Logic}. We demonstrate how a fuzzy characterization of NEAs can be obtained and how a subsequent analysis can deliver valid and quantitative results concerning the long-term dynamics of NEAs.

\end{abstract}

\section{Introduction and Motivation}
\label{intro}

\begin{figure}
\begin{center}
 \includegraphics[width=3.5in]{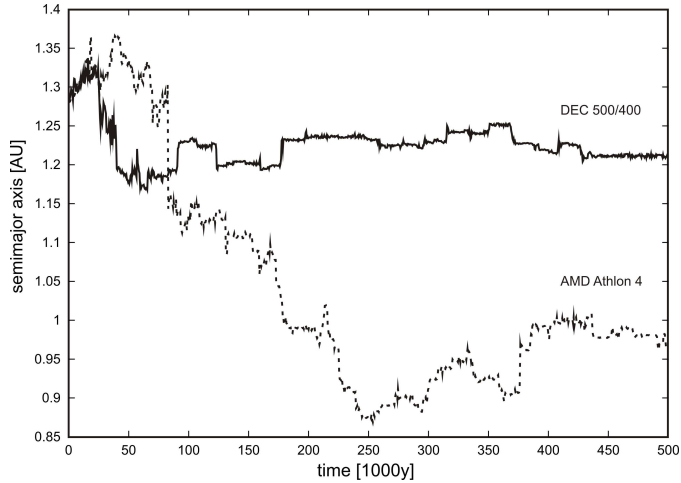}
\caption{Evolution of the semimajor axis of a NEA. The two curves were obtained with the same method and initial conditions but on different computers.}
\label{fig1}       
\end{center}
\end{figure}

Roughly between the orbits of Venus and Mars there exists a large amount of asteroids. Because they are able to cross the orbit of Earth they are called {\em Near-Earth-Asteroids (NEAs)}. Due to their proximity to Earth (and the other inner planets) they suffer from many close encounters during their dynamical evolution. For time scales longer than some ten thousand years those close encounters make the dynamics of NEAs strongly chaotic. Figure~\ref{fig1} shows an example for such a chaotic behavior. It depicts the evolution of the semimajor axis of a NEA over half a million years. The two curves were calculated with the same integration method using the same initial conditions. But we used two different computers with different processors and thus also different compilers. Therefore, very small differences were introduced and one can observe a classical feature of chaotic systems: tiny separations at the beginning will increase to large changes at the end of the integration time. In this case the two curves behave similar until the first close encounter of the asteroid with a planet (indicated by an abrupt change of the semimajor axis) and follow totally different trajectories afterwards. 

It is thus obvious that there will arise great difficulties in the investigation of single chaotic NEA orbits over long time scales. To overcome these problems one normally uses a statistical approach and studies groups of asteroids instead of single objects.\\
Currently there exist two major ways to classify NEAs into groups. Most widely used is the classification introduced by \cite{Shoemaker79}. In this scheme the NEAs are divided into three groups:

\begin{itemize}
\item the ATENS, with a semimajor axis smaller than the one of the 
Earth and an aphelion distance $Q = a (1+e) > 
0.983$ AU (mean perihelion distance of Earth)
\item the APOLLOS, with a semimajor axis larger than or equal 
to  the one of the Earth and a perihelion distance $q = a (1-e) 
\leq 1.017$ AU (mean aphelion distance of Earth)
\item the AMORS, with a semimajor axis larger than the one of the 
Earth and a perihelion distance $1.017 < q < 1.3$ AU 
\end{itemize}

Another classification was obtained during the {\em Project SPACEGUARD}~\cite{milani89} and is based on the dynamical properties of the asteroids using the following criteria:
\begin{itemize}
\item Values and changes of the orbital elements ($a,e,i,q,Q$)
\item Number and changes of node crossings (NC)
\item Number and depth of the close approaches (CA)
\item Resonances
\end{itemize}
According to these main criteria, one distinguishes between the following classes (for details see~\cite{milani89}): Geographos Class, Eros Class, Toro Class, Kozai Class, Alinda Class, Oljato Class and Comet Class.\\

However, if one is investigating the dynamics of NEAs for very long times then problems also arise using those two classifications. Due to the strong chaoticity of the motion, the classes do not hold anymore after a certain time. The phenomenon of asteroids that change their class during the investigated time scale was called {\em mixing} and already investigated in former works~\cite{freistetter04b,dvorak01,dvorak99}.\\
There it was found that an average NEA is a member of a certain group only for 69.86 \% of the integration time in case of Shoemakers classification and for 65.72 \% in the SPACEGUARD-classification. If asteroids change their initial group during the investigated time-span it is obvious that any statistical properties derived from this classification become ambiguous or invalid.\\
This problem is a general one and will occur for any classification simply based on orbital properties of NEAs. Due to their chaotic motion any group borders will be crossed after a certain time. Such classifications are thus only useful for sufficiently short time scales; if one is interested in longer times one has to use a totally different statistical approach.\\
In this work we intend to demonstrate that a new characterization of NEA dynamics can be obtained by means of {\em Fuzzy Logic}. Section~\ref{fuz} will thus give a short overview on the basic principles of fuzzy logic. Section~\ref{neas} will demonstrate how one can use methods from fuzzy logic to obtain a valid characterization of NEAs and section~\ref{exa} will give examples on how to apply this method.

\section{Fuzzy Characterization}
\label{fuz}
{\em Fuzzy set theory} or {\em Fuzzy Logic} was developed in 1965 by~\cite{zadeh65}. Fuzzy sets are an extension of classical sets. A
classical set is {\em
  two-valued}: for every set $A$ there exists a function $f_A$ that has either
the value $1$ or $0$ with:

\begin{equation}
f_A(x) = 1 \Leftrightarrow x \in A \hspace{0.1cm}\mathrm{and} \hspace{0.1cm} f_A(x) = 0 \Leftrightarrow x \notin A.
\end{equation}

\begin{figure}
\begin{center}
\includegraphics[width=3.in]{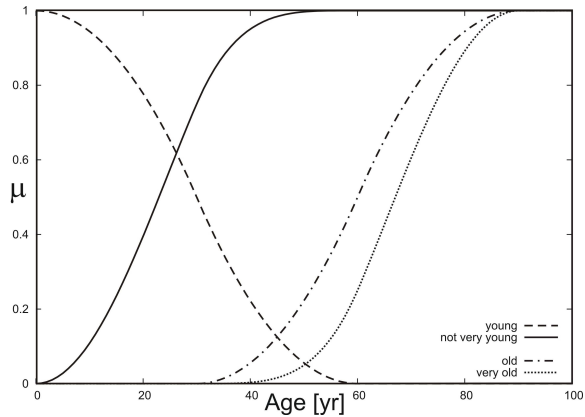}
\end{center}
\caption{Fuzzy membership functions for different sets describing the age of a person.}
\label{fuz1}
\end{figure}

This function is called {\em characteristic function} of $A$. Fuzzy sets, in
contrary, have a characteristic function $\mu_A$ defined for {\em all values
  between 0 and 1}, describing the degree\footnote{This must not be confused with a probability that an object belongs to a certain set!} to which an element $x$ is included
in the set $A$. Figure~\ref{fuz1} shows an example of the membership
functions describing the degree of membership to several classes related to the age of people. Here one can see one of the main advantages of fuzzy classification: objects can be members to different classes simultaneously with different degrees of membership: according to this functions a person of 30 years will have a degree of membership of $0.5$ to the group of {\em young persons}, of $0.75$ to the group of {\em not very young persons} and $0$ to the group of {\em old} and {\em very old persons}.\\

Fuzzy sets have the following properties:

\begin{itemize}
\item Classical sets can be interpreted as fuzzy sets with membership grades
  of only $0$ and~$1$
\item Two fuzzy sets $A$ and $B$ are equal over a whole set $X$ if
\begin{equation}
A = B \Leftrightarrow \mu_A(x) = \mu_B(x) \hspace{0.5cm} \forall x \in X
\end{equation}
\item For a fuzzy set A 
\begin{equation}
\label{alphacut}
A^{>\alpha} = \{x \in X \mid \mu_A(x) > \alpha \}, \hspace{1cm} \alpha \in [0,1]
\end{equation}
\begin{equation}
A^{\geq \alpha} = \{x \in X \mid \mu_A(x) \geq \alpha \}, \hspace{1cm} \alpha \in [0,1]
\end{equation}
are called the {\bf strong $\alpha$-cut} and the {\bf weak $\alpha$-cut},
respectively. 
\item
The $\alpha$-cuts of fuzzy sets are classical sets.
\end{itemize}

\section{Fuzzy Characterization of NEAs}
\label{neas}

The first step to obtain a fuzzy characterization for NEAs is a definition of fuzzy classes. There are many possibilities for a valid definition but to demonstrate the method we will restrict ourselves to 4 different classes connected to the planet-crossing behavior of the asteroids.\\
One of the most prominent features of the motion of NEAs is their ability to come very close to the inner planets. Thus it seems obvious to include this property into the definition of the fuzzy classes. We therefore define three groups for 
\begin{itemize}
\item Asteroids that can collide with Venus (G2)
\item Asteroids that can collide with Earth (G3)
\item Asteroids that can collide with Mars (G4)
\end{itemize}

The parameter that will be used to calculate the degree of membership for this classes is the number of close encounters with the respective planet: the more close encounters there are the larger the degree of membership will be.\\
Additionally we want to include a group that describes the {\em mixing} mentioned in section~\ref{intro}. Since the Shoemaker-classification depends only on the values of the semimajor axis $a$ and the eccentricity $e$ of an asteroid the mixing is here related directly to the changes of the orbital elements: asteroids that show no changes in $a$ and $e$ will also show no mixing; if however there are large changes in $a$ and/or $e$ the asteroid will almost certainly switch groups. We thus introduce a fourth group of

\begin{itemize}
\item Asteroids that show almost no mixing (G1).
\end{itemize}

This group allows us to monitor the behavior of semimajor axis and eccentricity and links the new fuzzy classification to the classical groups of Shoemaker. The parameter to calculate the degree of membership to this group will be a value called {\em Border Crossing Number (BCN)} defined as the number of times an asteroid crosses any group border in the Aten/Apollo/Amor classification. To calculate the number of close encounters and BCN we used an integration time of 500000 years.\\

In a next step the membership functions for the four groups have to be derived. To this purpose we investigate the distribution of the respective parameters (BCN and number of close approaches). We thus performed a numerical integration of the known NEAs using the Lie-Integration-Technique {see e.g.~\cite{Lichtenegger84,Hanslmeier84,asghari04}). Although the motion is not regularized this method is very accurate and also able to accurately threat close encounters between asteroids and planets (see e.g.~\cite{dvorak99,dvorak01}). The initial conditions of the NEAs were taken from the {\em JPL Horizons} database. We took into account the gravitational perturbations of Venus, Earth, Mars, Jupiter and Saturn. The integration was carried out for 500000 years and all close encounters with a distance smaller than 0.2 AU and all group changes of the asteroids in the Shoemaker-classification were recorded.\\
Figure~\ref{fig3} shows the resulting distribution of parameters.
\begin{figure}
\centerline{
\includegraphics[width=1.6in,angle=270]{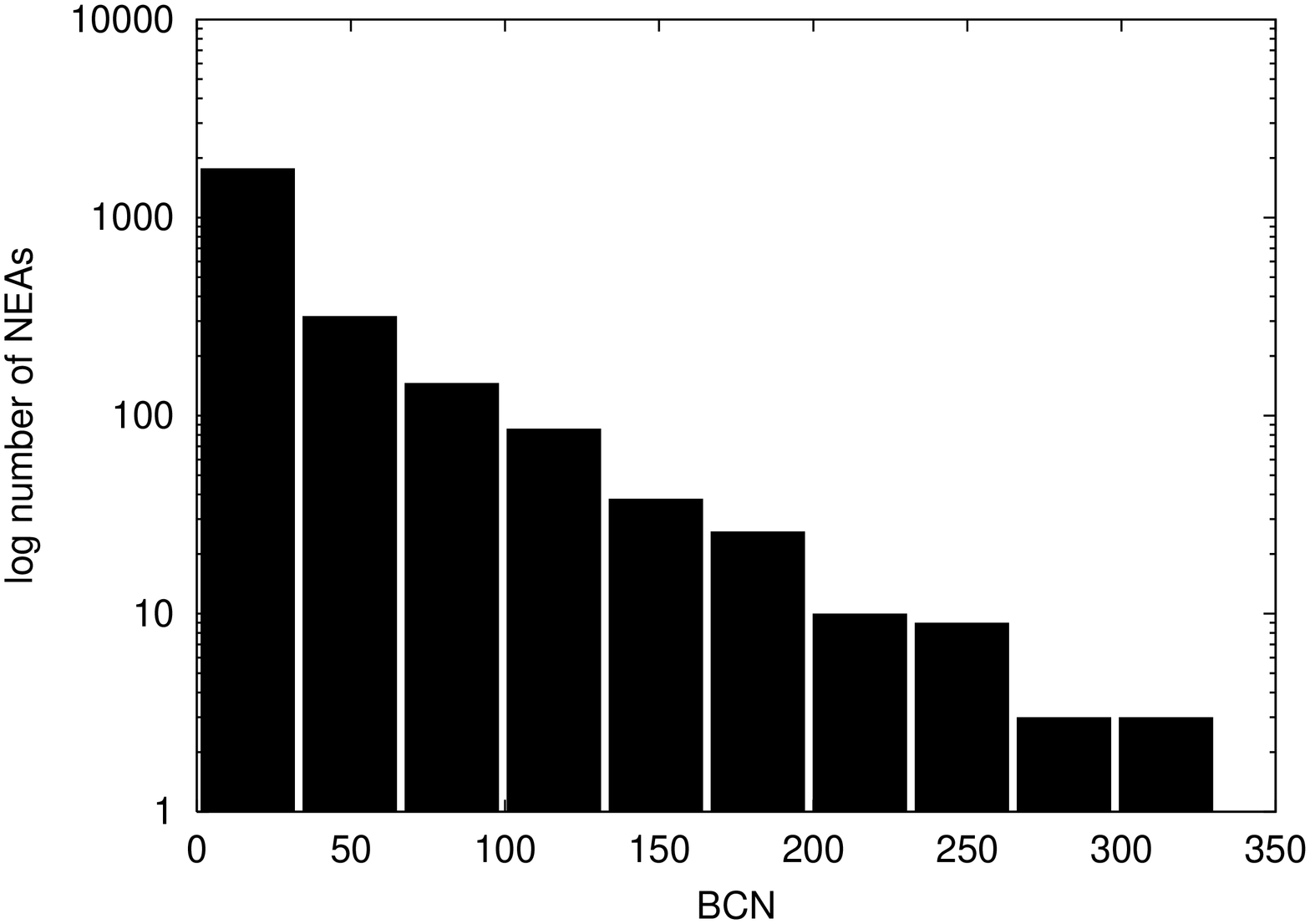}
\includegraphics[width=1.6in,angle=270]{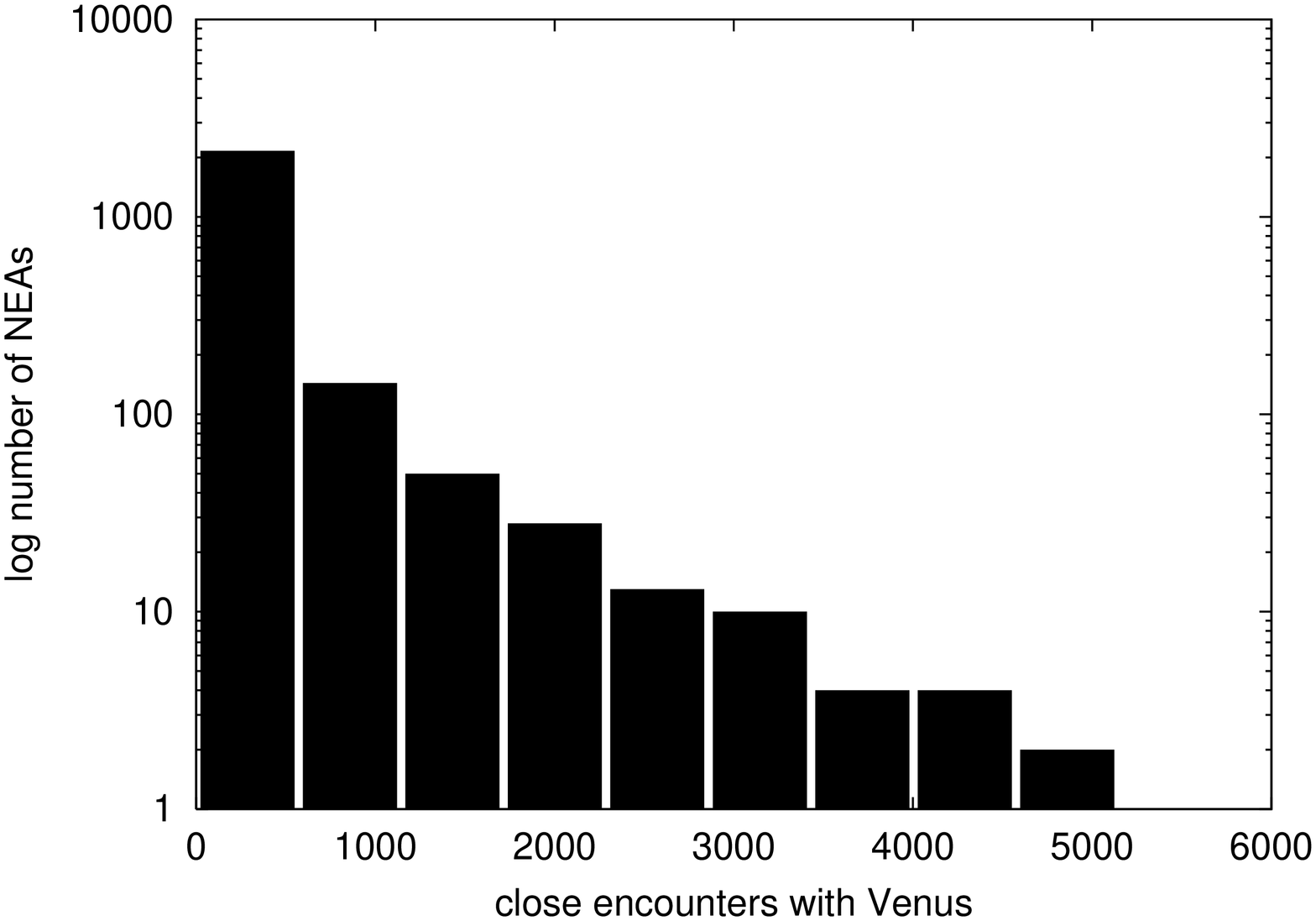}}
\centerline{
\includegraphics[width=1.6in,angle=270]{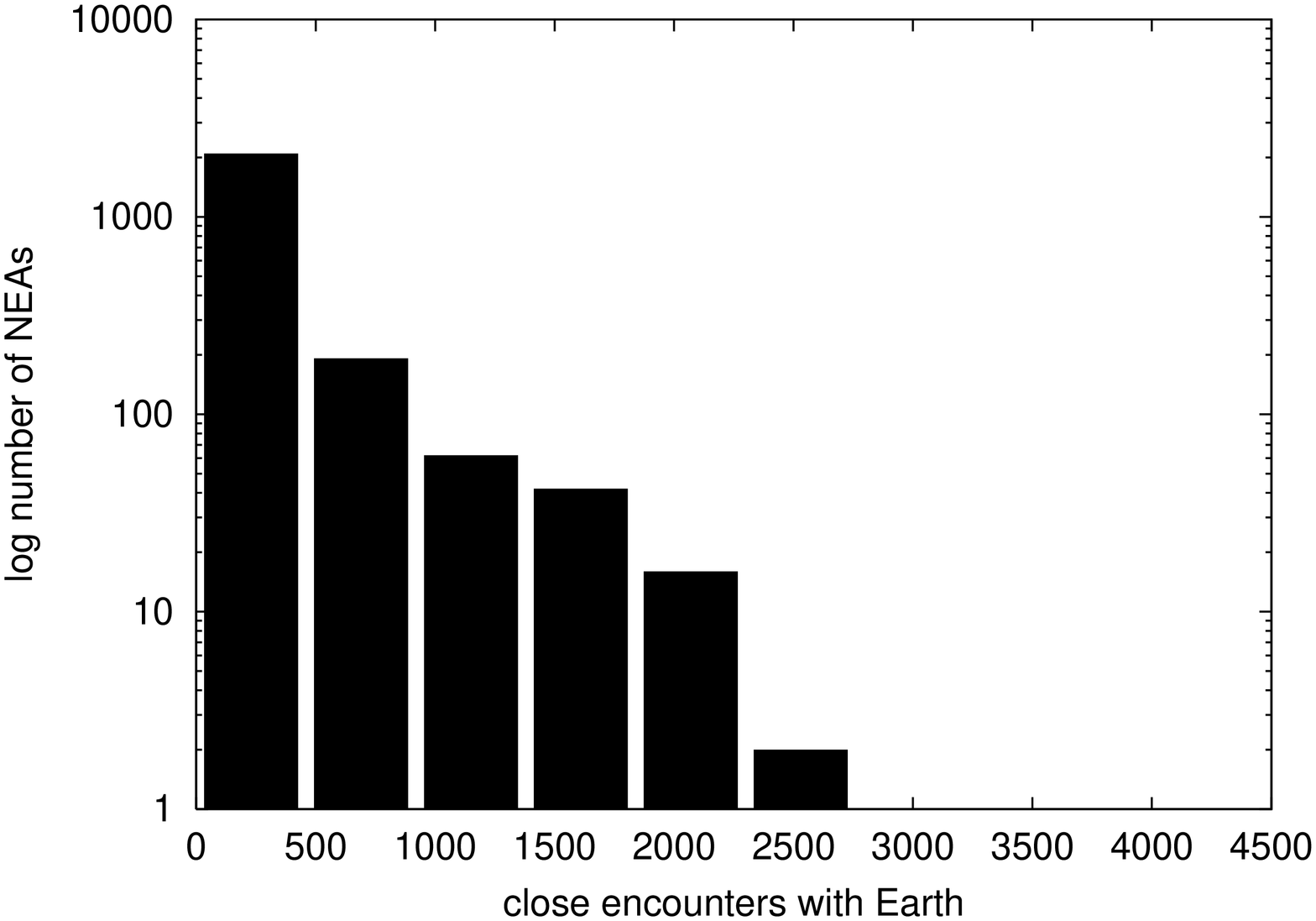}
\includegraphics[width=1.6in,angle=270]{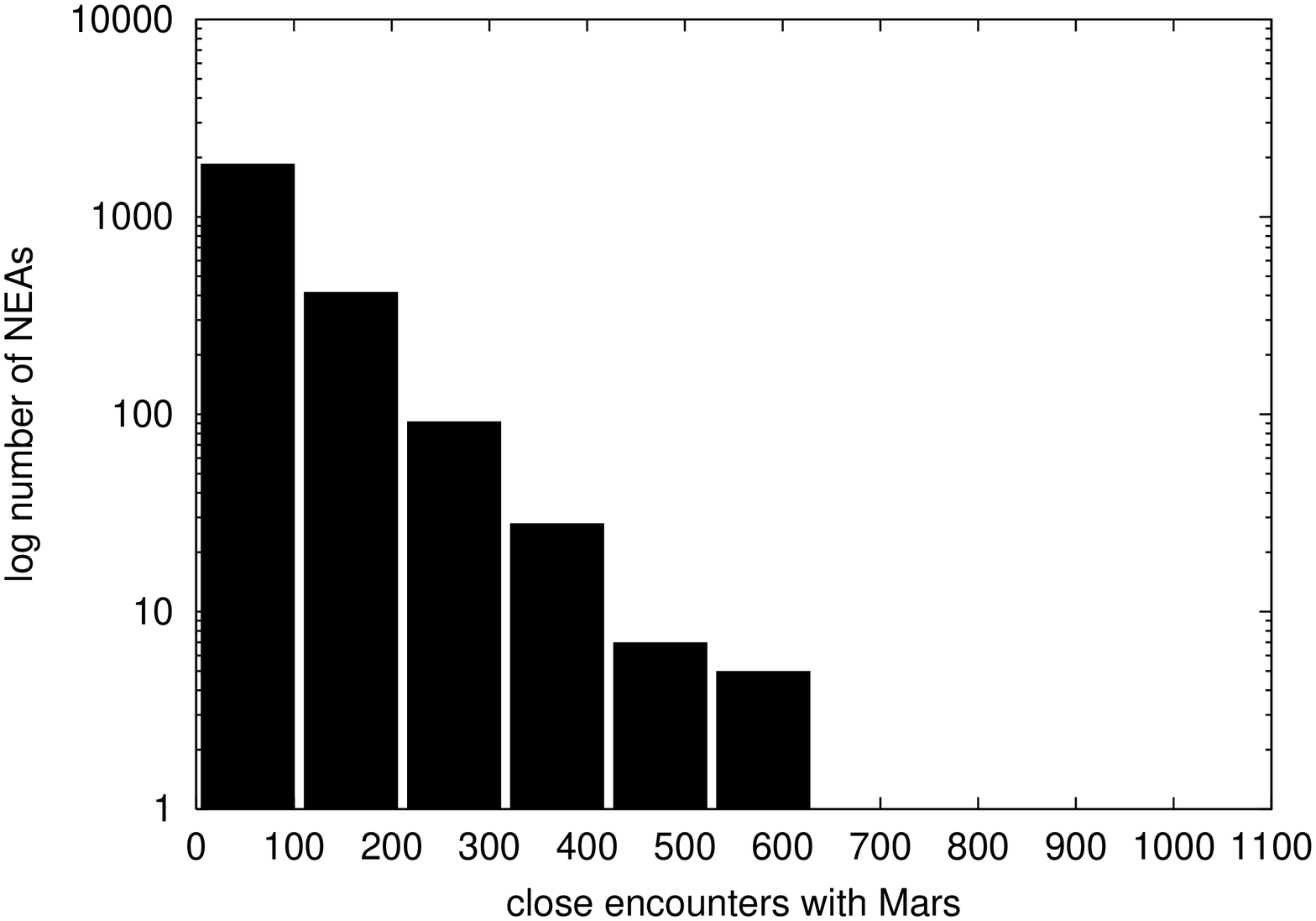}}
\caption{Distribution of BCN (top left) and close encounters with
  Venus (top right), Earth (bottom left), Mars (bottom right)Note that the y-scale is logarithmic.}
\label{fig3}
\end{figure}
This distributions were used to derive the membership functions. We fitted a function through the histogram data and normalized it so that it delivers only values between $0$ and $1$. In case of close encounters, the functions additionally have to be mirrored to reflect the correct behavior: if there are no close encounters, the degree of membership should be zero and the more close encounters there are the larger the degree of membership has to be. Figure~\ref{fig4} shows the final membership functions for the four classes.\\

\begin{figure}
\centerline{
\includegraphics[width=1.6in,angle=270]{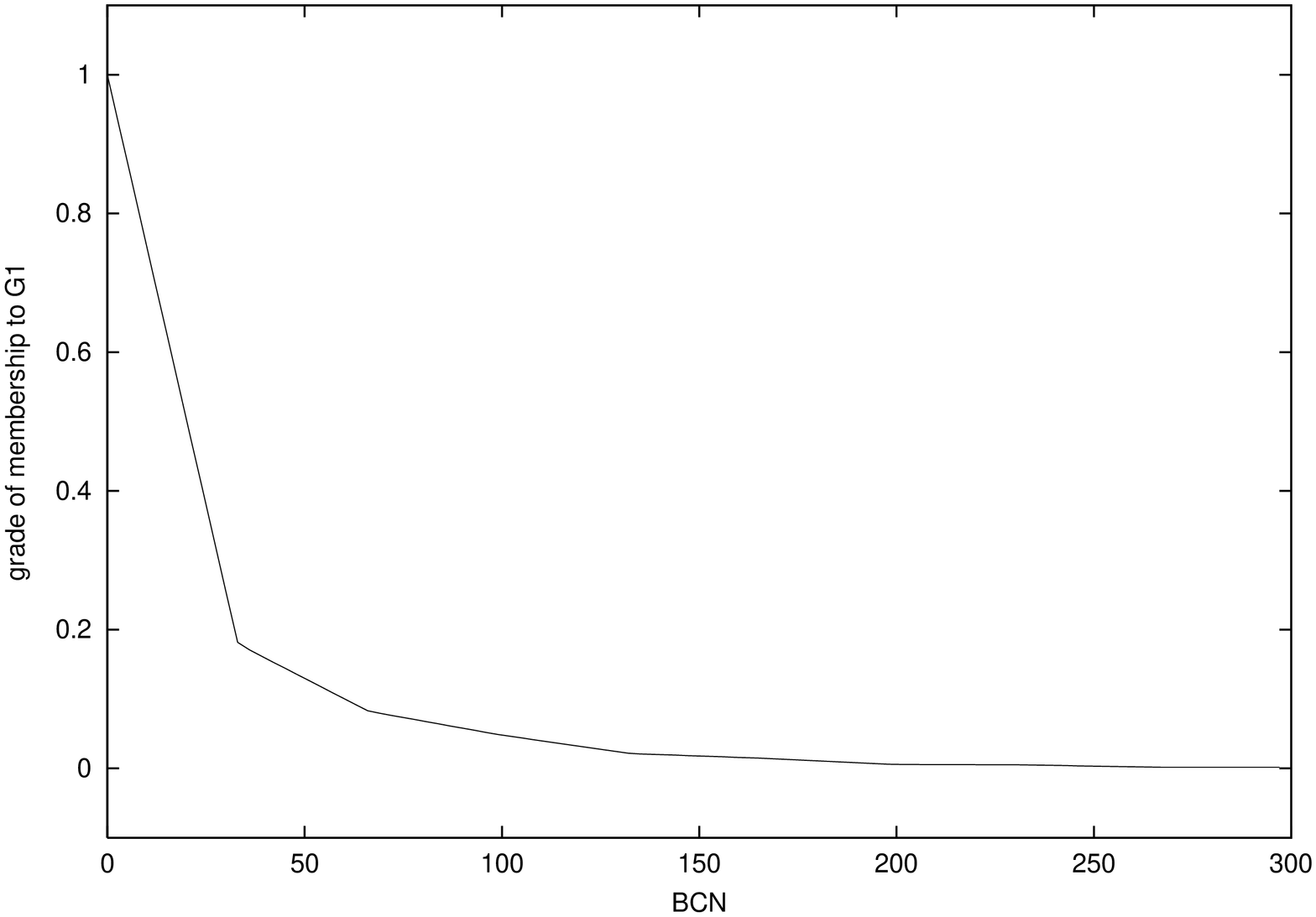}
\includegraphics[width=1.6in,angle=270]{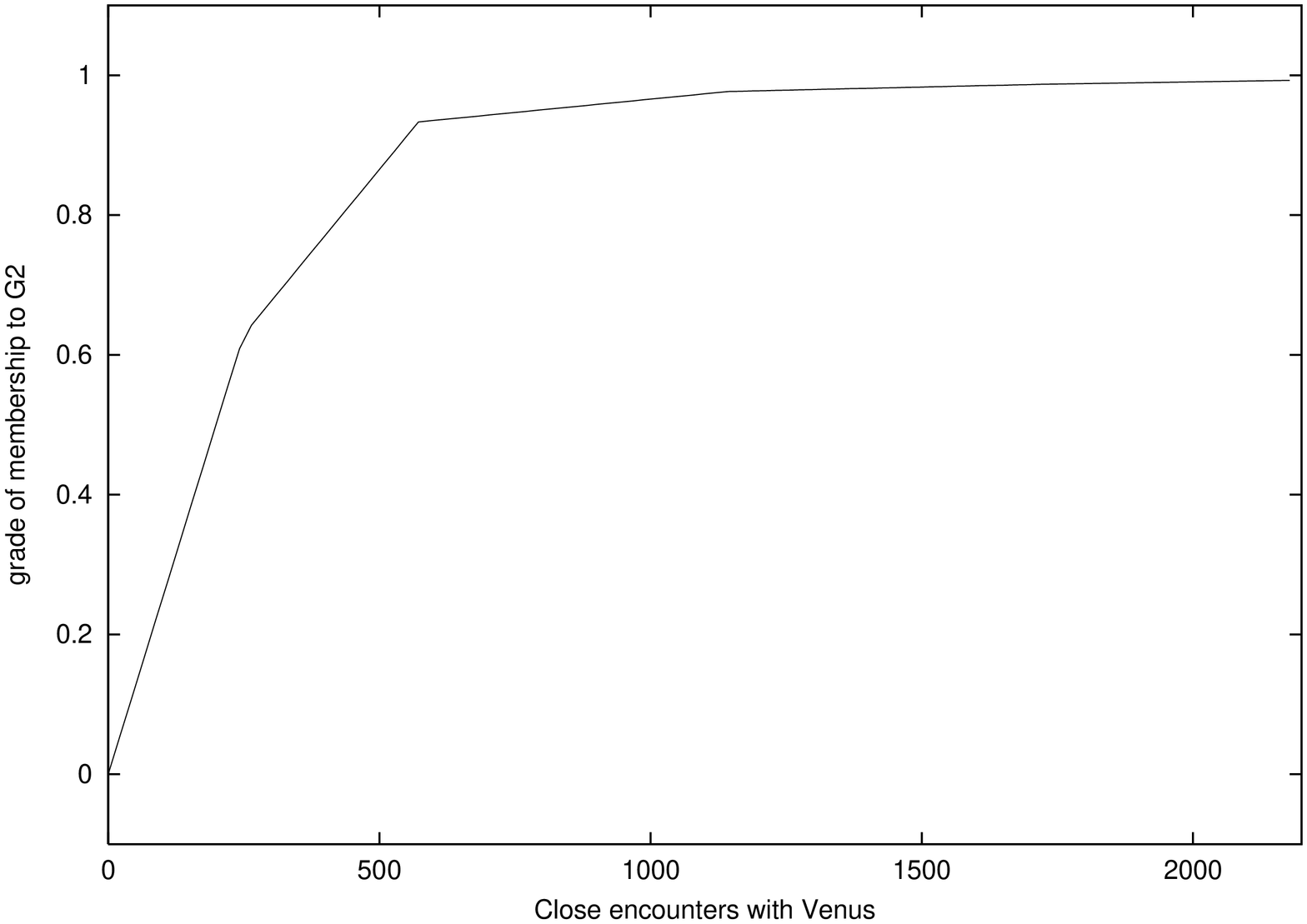}}
\centerline{
\includegraphics[width=1.6in,angle=270]{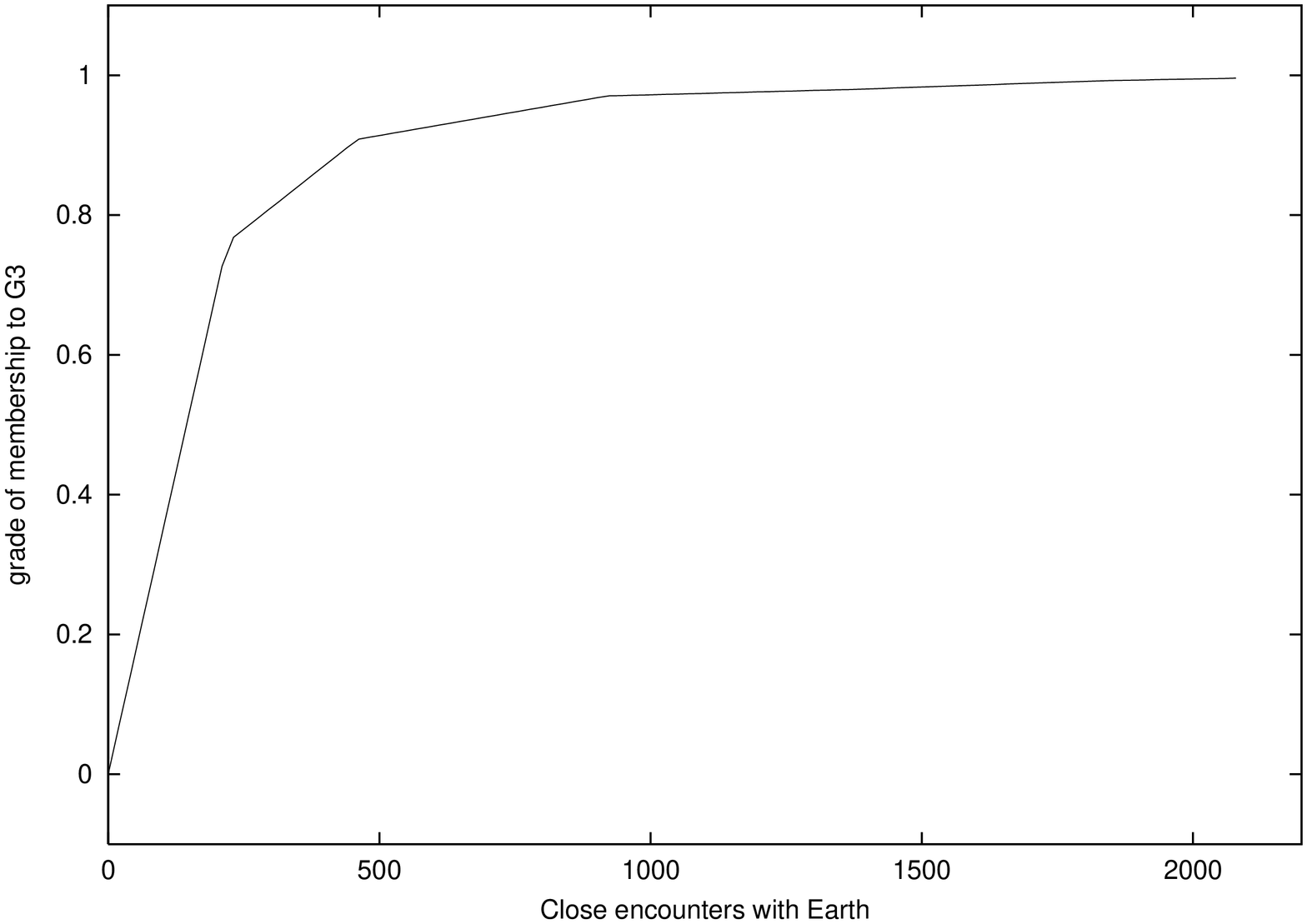}
\includegraphics[width=1.6in,angle=270]{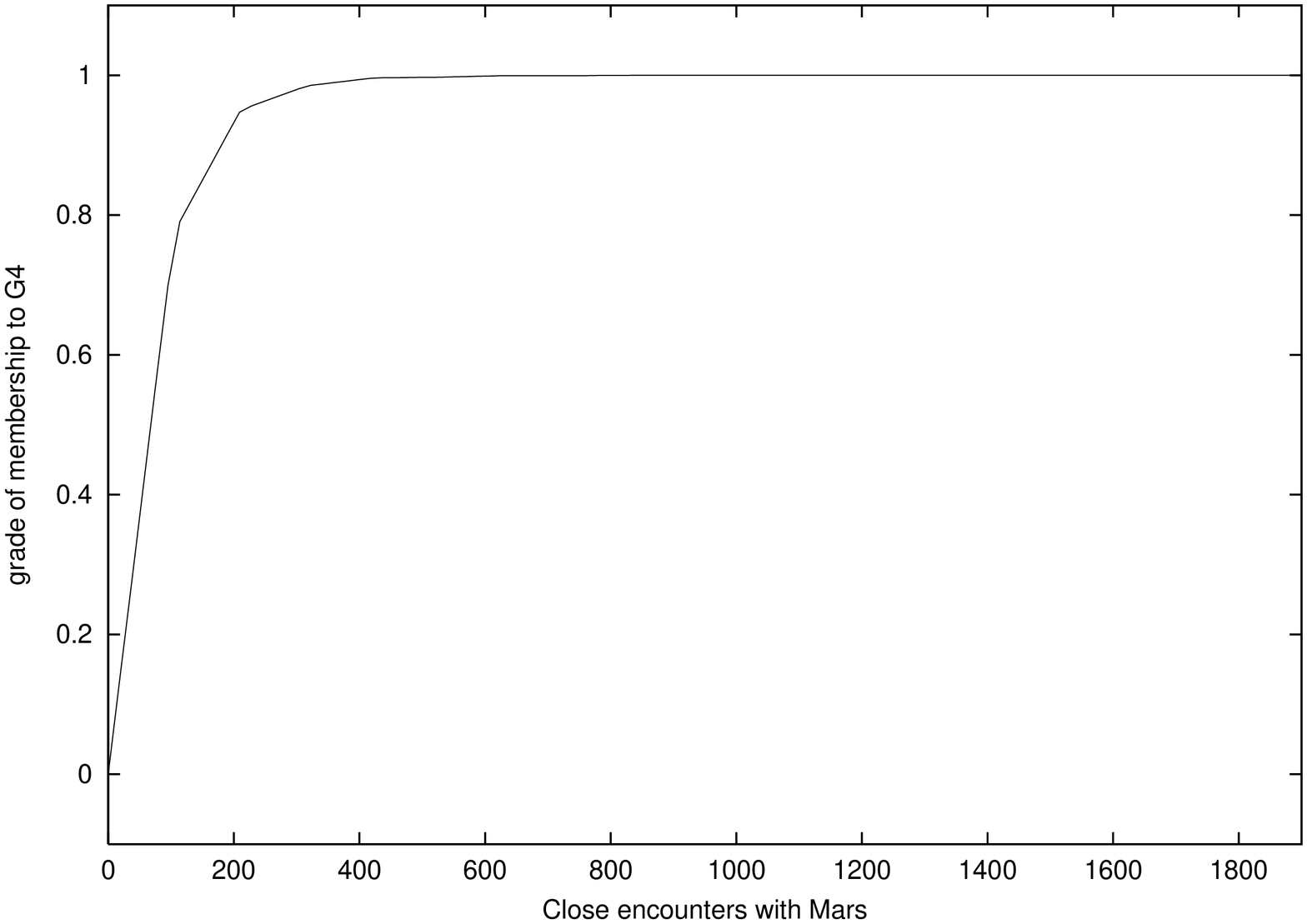}}
\caption{Membership function for the group G1 (top left), G2 (top right), G3 (bottom left), G4 (bottom right).}
\label{fig4}
\end{figure}

The membership for the classes G2, G3 and G4 are very similar. This was to be expected because we deal with the long-term evolution of chaotic asteroids. An object that has a close encounter with one planet has a large probability to also have close encounters with the other planets.\\

We now are ready to classify the asteroids: using the membership functions a degree of membership to each of the four groups can be calculated for every asteroid. This was done for all NEAs (the numerical values can be obtained online at {\tt http://celestialmechanics.eu/neas}). In a strict sense the membership functions are only valid for the given integration time. Because the motion of NEAs is chaotic, one can not predict the orbital motion and the number of close encounters and thus the distributions and functions shown in figures~\ref{fig3} and~\ref{fig4} can change if the motion of the asteroids is calculated for longer times.

Before we will demonstrate how one can use the fuzzy classification to analyze the dynamics of NEAs we will first verify that this fuzzy characterization indeed is possible to describe the motion NEAs correctly. To this purpose we compare the fuzzy groups with the SPACEGUARD-classification.  As~\cite{milani89} have also classified the asteroids according to their collision probabilities and close encounters the results should be consistent -- at least there, where the two classifications are comparable. For comparison, we will look at the namesakes of the seven SPACEGUARD classes (which, according to Milani et. al. are the most prominent examples of their classes):

 \begin{itemize}

\item{\bf (1620) Geographos: } according to the SPACEGUARD classification, an
  asteroid of the Geographos group should show many close approaches to Earth
  and some to Venus. The semimajor axis of a Geographos is
  almost constant; the eccentricity shows secular trends on small scales --
  thus it is not expected to move very much in the $a-e$ plane.\\
 (1620) Geographos has a membership grade to G1 of $1$ -- so it is indeed a full member of the group of asteroids that show almost no mixing and its semimajor axis and
  eccentricity are not expected to change very much. Also the membership grade to G2
  (0.03) and G3 (0.76) reflect the behavior described in the SPACEGUARD classification. The membership
  grade to G4 (0.86) shows now the influence of the longer integration time: during
  the orbital eviolution of the asteroid, some deep close encounters can change drastically the
  semimajor axis of an asteroid inside the Geographos class and thus it into the vicinity of Mars. 
Another object in the Geographos group to demonstrate how the problems
  of mixing are bypassed by the new fuzzy classification is (1862) Apollo: in the SPACEGUARD
  classification, after a close encounter with Venus, (1862) Apollo was no
  longer a member of the Geographos group. In the fuzzy classification,
  however, all the dynamical properties of all groups are described simultaneously.

\item{\bf (1685) Toro: } according to the SPACEGUARD classification an asteroid
  of the Toro group shows close approaches with Earth. In general, the semimajor axis and eccentricity show only small variations.\\
  (1685) Toro indeed has a membership grade to G1 of $1$ -- so its semimajor
  axis and eccentricity  are not expected to change very much. The membership
  grade to G2 (asteroids that can collide with Venus) is $0$ --
  also in~\cite{milani89} (1685) Toro is in resonance with Venus and thus
  protected from close encounters. Grade of Membership to G3 is 0.77 which corresponds to many close approaches with Earth, as expected. 

\item{\bf (1863) Antinous: } it was not possible to compare the results for
  (3040) Kozai (the most prominent member of the class of Kozai asteroids)
  because although it belongs to the Mars-crossing asteroids, its perihelion
  distance is larger than $1.3$ and thus is not a NEA in strict sense (Milani
  et al. did not just use NEAs but all planet crossing asteroids for their
  classification). We thus use (1863) Antinous for our comparison which is also a member of the Kozai group. Kozai asteroids are, due to Kozai resonances~\cite{kozai62, kinoshita07}
  of type I, protected against close encounters and collisions. The evolution
  of the semimajor axis is very regular and shows only small oscillations. The
  group of Kozai asteroids is the most stable class in the SPACEGUARD
  classification.\\ (1863) Antinous
  indeed shows the described behavior: the grade of membership to G1 is $1$,
  that to G2 (0.23), G3 (0.16) and G4 (0.14) is considerably smaller than that
  of the fore mentioned asteroids which corresponds to very few close encounters and reflects the stability of the group.

\item{\bf (887) Alinda: } according to SPACEGUARD classification, asteroids of
  the Alinda group are in (low order) mean motion resonances with Jupiter. Their eccentricities can undergo large changes, the semimajor axes
  oscillate around the resonant value. Because of the probable large changes
  asteroids can encounter all inner planets but are often protected against
  collision by resonances. (887) Alinda
  has a smaller grade of membership to G1 (0.93) than the asteroids mentioned
  before indicating the larger changes in $a$ and $e$; also the grades of membership to G2 (0.12), G3 (0.08) and G4 (0.04)
  are considerably smaller -- showing the resonant protection.

\item{\bf (433) Eros: } according to SPACEGUARD classification, asteroids of
  the Eros group are those, which do not cross the orbit of Earth because
  their perihelion is always higher than $1$ AU. All Eros asteroids are Mars
  crossers and have close approaches with Mars. The eccentricities of Eros asteroids can
  show very large changes. Indeed the grade of
  membership of (433) Eros to G2 and G3 is $0$, it is only a member of G4
  (0.72). Also the membership to G1 (0.9) is smaller than 1, indicating the
  larger changes of $a$ and $e$. Another good example for the behavior of Eros
  asteroids is
  (719) Albert. Its membership to G1 (0.08) is very low (indicating very large
  changes in $a$ and $e$), again the grade of
  membership to G2 and G3 is zero and that to G4 is 0.32.

\item{\bf (2201) Oljato: } according to the SPACEGUARD classification, asteroids of
  the Oljato group have orbits that show large-scale chaotic effects. They
  have very high eccentricities and can have close approaches to all inner
  planets.  (2201) Oljato indeed has a grade
  of membership to G1 of 0.04, indicating the chaotic changes in $a$ and $e$;
  it also shows a medium grade of membership to G2 (0.38), G3 (0.36) and G4
  (0.28) -- thus it encounters all inner planets.

\item{\bf Comet class: } a comparison with the class of Comet asteroids of the
  SPACEGUARD classification is not possible in this work. This class consists
  of all asteroids that spent a sufficient part of integration time in the
  outer solar system which were excluded from our study.
\end{itemize}

The comparison shows that the fuzzy characterization is indeed capable to describe the dynamics of NEAs correctly. In contrary to the SPACEGUARD classification, the new fuzzy classification now takes into account all dynamical properties simultaneously. Thus no group changes can occur which bypasses the problem of mixing.

\section{Examples of Analysis}
\label{exa}
The analysis of fuzzy groups is based on the concept of $\alpha$-cuts introduced in section~\ref{fuz}. An $\alpha$-cut allows one to extract a classical set out of a fuzzy set which is then available to analysis. \\

\subsection{Connection between the planet-crossing populations}

This will be demonstrated with the set $G3^{>0.9}$ that contains all NEAs with a degree of membership to the group G3 (``asteroids that can collide with Earth'') larger than 0.9. We now have a group of objects with a common major property (the degree of membership to G3 is larger than 0.9) but different additional properties (their degree of membership to the other groups) and can analyze them accordingly.\\
Figure~\ref{fig5} shows the distribution of those asteroids according to their grade of membership\footnote{Here and in the following,
  {\em high} grades of membership are defined by $\mu_{Gi} \ge 0.9$,
  {\em medium} grades by $0.1 \le \mu_{Gi} < 0.9$ and {\em low} grades by
  $\mu_{Gi} < 0.1$.} to G2 and G4.

\begin{figure}
\centerline{
\includegraphics[width=1.6in,angle=270]{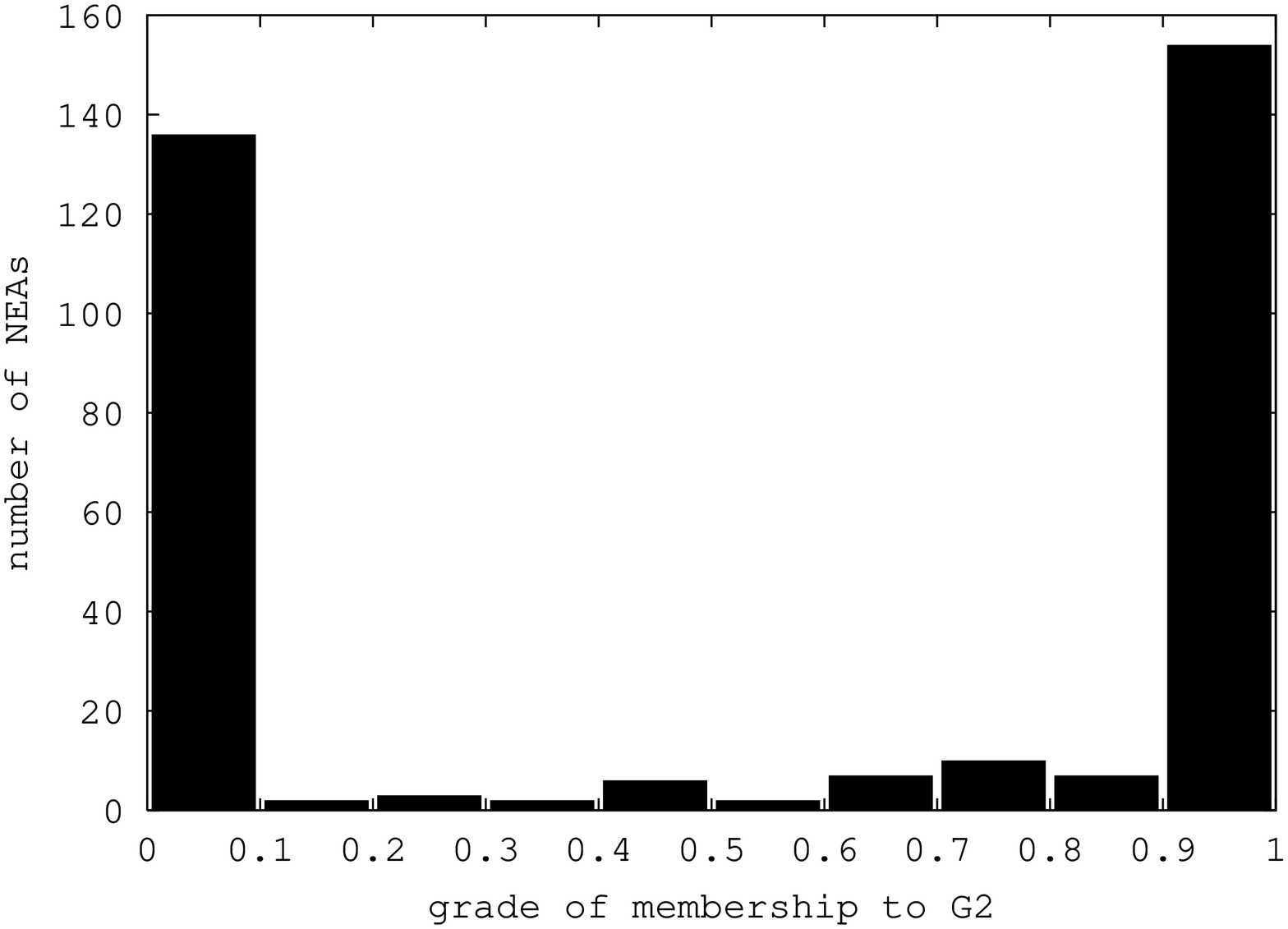}
\includegraphics[width=1.6in,angle=270]{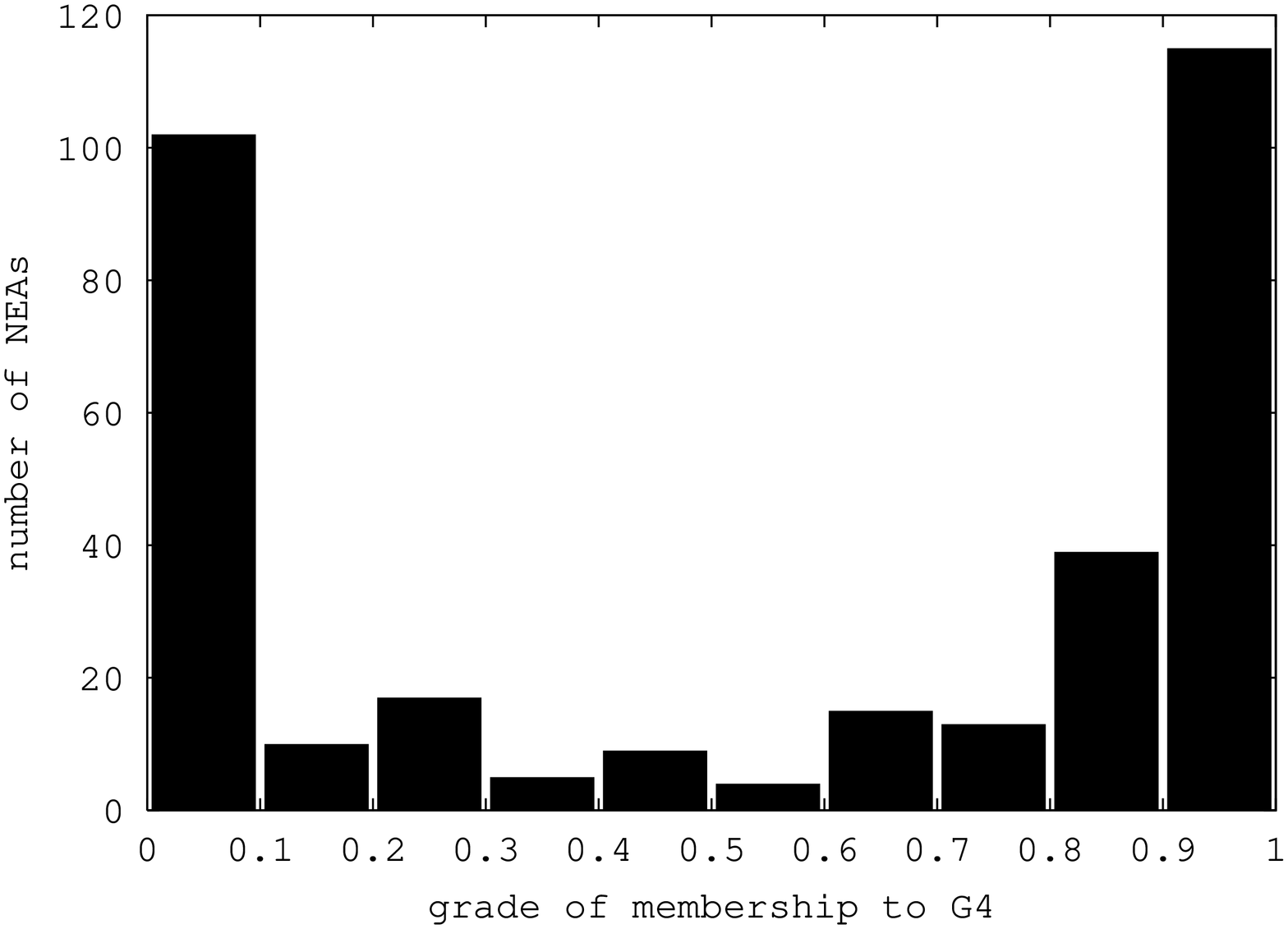}}
\caption{Distribution of asteroids in the group  ${G3}^{ >0.9}$  according to their grade of membership to G2 and G4.}
\label{fig5}
\end{figure}

When looking on the medium grades of membership to G2 and G4, the figure immediately shows an important feature:  only $11.85\%$
of the asteroids have medium grades to G2, whereas three times more of
them ($34.04\%$) have medium grades to G4. Thus, the
``connection'' between Earth and Mars-crossing asteroids is more
fluent: NEAs that are likely to collide with Earth in the majority are
also likely to collide with Mars -- and, also in the majority, are
likely to collide with Venus (this can be seen by the large amount of objects with high grades of membership to G2 and G4); but the lack of intermediate grades of
membership to G2 shows that the interaction between Earth and Venus is
much stronger. {\em If} deep close encounters bring an asteroid near
Venus (which is the case for slightly more than half of asteroids), it
is very probable that they have very much close encounters (and thus
also a higher collision probability) with Venus. On the other hand, if
they come close to Mars (which is also the case for slightly more than
half of asteroids) the probability that they have a high or
intermediate number of close encounters is almost equal ($34.04\%$ of
$G3^{>0.9}$ have medium grades of membership, $34.95\%$ have high grades). Earth is able to ``protect'' its crossing
asteroids much more easily from the influence of Mars than that of Venus (which is certainly due to the smaller mass of Mars compared to that of Venus).\\

This example has demonstrated how a quantitative analysis of the chaotic long-term behavior can be obtained when one uses a fuzzy characterization. A second example will give additional insights in the usefulness of the method.\\

\subsection{Long-term fate of NEAs}

NEAs are an unstable population of celestial bodies. Due to their chaotic motion they will end their life in a collision with a planet, in a collision with the sun or in an ejection from the solar system. The fuzzy classification can help to shed light on the fate of NEAs. As exemplified above, the long-term motion of NEAs is governed by close encounters with the inner planets. If viewed over some hundred thousand years, an orbit of a NEA is thus ``smeared'' over the whole inner solar system and the asteroid will visit every region between the orbits of the inner planets. Since Venus orbits the sun faster than Earth also the probability of a collision with Venus should be larger than with the Earth for the average NEA. Indeed, the ratio of mean-motions of Venus and Earth is $1.38$ which is very close to the ratio of collision probabilities of NEAs with Venus and Earth (see e.g.~\cite{dvorak99,freistetter04b}).\\
A fuzzy characterization and an according analysis can help to obtain a different view on the long-term fate of NEAs.
As in the first example, we build three $\alpha$-cuts: $G2^{>0.9}$, $G3^{>0.9}$ and $G4^{>0.9}$. Those sets of objects contain all NEAs with a common major property: a high degree of membership to the class of asteroids that can collide with Venus (Earth, Mars). In this case the three $\alpha$-cuts are not mutually exclusive: an asteroid can be member of two or even all three $\alpha$-cuts. We thus investigated the mutual membership of the three classes. The results are represented in figure~\ref{fig6}.

\begin{figure}
\centerline{
\includegraphics[width=3.2in,angle=0]{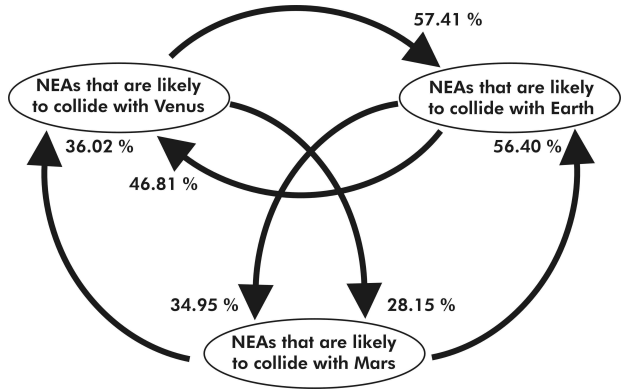}}
\caption{Groups of asteroids that are likely to collide with a planet. The
  arrows show, how many NEAs of one group, are also members of an other group.}
\label{fig6}
\end{figure}

$G3^{>0.9}$ (corresponding to Earth) is not only itself 
the largest of all three groups, also the asteroids in the other groups are more often also members in $G3^{>0.9}$ than
vice versa! This leads to the following conclusion:
NEAs move on orbits with semimajor axes from $\sim 0.6$ to $\sim 3$ AU (depending on their eccentricity). They can come close
(and also collide) with all large inner planets. As long as they are NEAs (and not have collided with another body or have been ejected), independently from their initial position in
the $a-e$ plane, they
have the tendency to evolve Earth-crossing orbits. Thus, for time scales comparable to the integration time of this work, the major reason for the decrease of NEA population will be due to collisions with Earth.

\section{Conclusions}
\label{con}

We have demonstrated that the chaotic motion of NEAs makes a statistical treatment impossible or at least very difficult. We introduced methods from fuzzy logic to overcome these problems and showed that the resulting fuzzy characterization of NEAs gives valid outcomes and can incorporate the existing classifications. Additionally, we exemplified that a fuzzy characterization allows a quantitative statistical analysis of NEA dynamics.\\

This work only displays the first step; additional work will give a more detailed and practical fuzzy characterization of NEA dynamics. The number of known NEAs increases almost every day and the new orbital data has to be incorporated to refine the membership functions. Instead of the current group G1 that is based on the mixing behavior it could be an improvement to directly include the variation of $a$, $e$ and also the inclination of the asteroid orbits as basis of fuzzy class. Forthcoming studies will also extend the integration time to obtain a better representation of the long-term behavior of NEAs. To describe the motion of NEAs in more detail it is also advisable to introduce additional fuzzy classes that e.g. describe the resonant behavior of the objects.\\
Future work will be dedicated to tackle this problems and deliver an accurate description of the long-term motion of NEAs.


%




%
%

\end{document}